\documentclass[useAMS,usenatbib]{mn2e}

\title[Early BHs in Cosmological Simulations]{Early Black Holes in Cosmological Simulations: Luminosity Functions and
  Clustering Behaviour}

\author[Colin Degraf et al.]  {Colin DeGraf$^{1}$,
       Tiziana Di Matteo$^{1}$, Nishikanta Khandai$^{1}$, Rupert Croft$^{1}$,
       \newauthor
       Julio Lopez$^{2}$, Volker Springel$^{3,4}$\\ 
       $^{1}$ McWilliams Center for
       Cosmology, Carnegie Mellon University, 5000 Forbes Avenue, Pittsburgh,
       PA 15213, USA\\
       $^{2}$ Computer Science Department, Carnegie Mellon University, 5000
       Forbes Avenue, Pittsburgh PA 15213, USA\\
       $^{3}$ Heidelberg Institute for Theoretical Studies,
       Schloss-Wolfsbrunnenweg 35, 68118 Heidelberg, Germany\\
       $^{4}$ Zentrum f\"{u}r Astronomie der
Universit\"{a}t Heidelberg, Astronomisches
Recheninstitut, M\"{o}nchhofstr. 12-14, 69120 Heidelberg, Germany
}

\usepackage{ graphicx}
\usepackage{ subfigure}
\usepackage{multirow}
\usepackage{hyperref}

\def\simgt{\lower.5ex\hbox{$\; \buildrel > \over \sim \;$}}

\begin{document}

\date{Submitted to MNRAS}
\pagerange{\pageref{firstpage}--\pageref{lastpage}}
\pubyear{20??}

\maketitle
\begin{abstract}
We examine predictions for the quasar luminosity functions (QLF) and quasar
clustering at high redshift ($z \ge 4.75$) using {\it MassiveBlack}, our new hydrodynamic
cosmological simulation which includes a self-consistent
model for black hole growth and feedback.  We show that the model
reproduces the Sloan QLF within observational constraints at $z \ge 5$.  We
find that the high-z QLF is consistent with a
redshift-independent occupation distribution of BHs among dark matter halos
(which we provide)
such that the evolution of the QLF follows that of the halo mass function.  The sole
exception is the bright-end at $z=6$ and 7, where BHs in
high-mass halos tend to be unusually bright due to extended periods of
Eddington growth caused by high density cold flows into the halo center.  We
further use these luminosity functions to make predictions for the number density of quasars in upcoming surveys,
predicting there should be $\sim 119 \pm 28$ ($\sim 87 \pm 28$) quasars
detectable in the F125W band of the WIDE (DEEP)
fields of the Cosmic Assembly Near-infrared Deep Extragalactic Legacy Survey
(CANDELS) from $z=5-6$, $\sim 19 \pm 7$ ($\sim 18 \pm 9$) from
$z=6-7$, and $\sim 1.7 \pm 1.5$ ($\sim 1.5 \pm 1.5$) from $z=7-8$.
We also investigate quasar clustering, finding that the correlation length is fully consistent with current
constraints for Sloan quasars ($r_0 \sim 17 \: h^{-1} \: \rm{Mpc}$ at $z=4$ for quasars above
$m_i = 20.2$), and grows slowly with redshift up to $z=6$ ($r_0 \sim 22 \: h^{-1}
\: \rm{Mpc}$).  Finally, we note that the quasar clustering strength depends weakly on luminosity for
low $L_{\rm{BH}}$, but gets stronger at higher $L_{\rm{BH}}$ as the BHs are
found in higher mass halos. 

%Finally, we note
%that the unusually bright BHs undergoing Eddington growth should cause a minor
%suppression in clustering strength at $z \sim 6-7$.  

%We use a new, extremely large hydrodynamic cosmological simulation to probe
%the statistical properties of high redshift ($z \ge 4.75$) black holes.  We
%compute the Quasar Luminosity Function (QLF) up to z=11, finding reasonable
%agreement with current observations and making predictions for the numbers of
%quasars which should be found by several upcoming surveys, noting that they
%will drastically improve the constraints on the faint-end QLF slope.  We find that the
%high-z Luminosity Function is consistent with a redshift-independent
%occupation distribution of BHs among dark matter halos, except at z=6 \& 7,
%where BHs in high-mass halos tend to be unusually bright due to extended
%periods of Eddington growth.  We also compute the clustering strength of high-z quasars, finding correlation lengths consistent with current observational constraints but extended
%to much higher redshift.  We show
%that the clustering strength depends weakly on $L_{\rm{BH}}$ for low
%luminosity, but gets stronger at higher $L_{\rm{BH}}$.  Finally, we note that
%the unusually bright BHs undergoing Eddington growth should
%cause a minor suppression in clustering strength at $z \sim 6-7$.
\end{abstract}


\section{Introduction}

Supermassive black holes are believed to be present at the center of most
galaxies \citep{KormendyRichstone1995} and are found to be correlated with the properties
of their host galaxies \citep{Magorrian1998, FerrareseMerritt2000,
  Gebhardt2000, Tremaine2002, GrahamDriver2007}.  These correlations provide
strong evidence for a link between the growth of black holes and the evolution
of their host galaxies, generally attributed to some form of quasar feedback
\citep{BurkertSilk2001, Granato2004, Sazonov2004, Springel2005, Churazov2005,
  KawataGibson2005, DiMatteo2005, Bower2006, Begelman2006, Croton2006,
  Malbon2007, CiottiOstriker2007, Sijacki2007, Hopkins2007}.

Perhaps the most fundamental statistical quantity in the study of quasars is the number
density, often characterized as a Quasar Luminosity Function (QLF).  The QLF
has been studied observationally \citep{LaFranca2002, Fiore2003, Ueda2003,
  Barger2003, Croom2004, LaFranca2005, Cirasuolo2005, Richards2006, Brown2006,
  Silverman2008, Ebrero2009, Yencho2009}, as well as through simulations, with
black holes modeled both
semi-analytically \citep{KauffmannHaehnelt2000, Volonteri2003, WyitheLoeb2003,
Granato2004, Marulli2008, Bonoli2009} and directly incorporated in the
simulation \citep{Hopkins2005,
  HopkinsQLF2006, Hopkins2007, Marulli2009, DeGraf2010}.  

Another fundamental quantity for quasar populations is the strength of quasar
clustering, and its evolution with redshift.  Numerous observational studies
have been done \citep{LaFranca1998, Porciani2004, Croom2005, Shen2007,
  Myers2007, daAngela2008, Shen2009, Ross2009}, generally finding evidence for
increasing clustering amplitude with redshift, in agreement with findings from
simulations \citep{Bonoli2009, Croton2009, DeGraf2010}.  Clustering is
especially significant because one can use it to estimate the typical dark matter
halos in which quasars are found simply by comparing the
clustering strength of quasars to that of dark matter halos. In particular,
the luminosity dependence of the clustering (if any) can help determine if the
bright and faint quasars populate the same halos, suggesting they may be the
same population of quasars at different phases of their lives \citep[see, e.g.][]{Hopkins2005,
  Hopkins2005b, Hopkins2005c, Hopkins2005d, Hopkins2006b} or
different halo masses, suggesting they are fundamentally different quasar
populations \citep[see, e.g.][]{Lidz2006}.
%, thereby providing a means of understanding how $L_{\rm{BH}}$ correlates with host halo mass.

Thus both QLF and quasar clustering can be used to investigate the populations
of quasars being observed, how they populate dark matter halos, and how the
typical luminosities correlate with hosts.  However, investigations at high redshift
have been extremely difficult.  Simulations have proved difficult due to the
volumes needed to produce high redshift quasars in sufficient numbers,
limiting the statistical investigations.  Similarly, the difficulty observing such
distant objects limits the current number of observed high redshift quasars such that the $z \sim 6$ faint-end QLF is determined by only a
few objects \citep{WillottCFHQS2010}, and high-redshift quasar clustering is
limited to lower redshift \citep[e.g. $z \sim 4$ by][]{Shen2009}.  However, upcoming
surveys such as the James Webb Space Telescope (JWST) and the Cosmic Assembly
Near-infrared Deep Extragalactic Legacy Survey (CANDELS) have the potential to
drastically improve the high-z observations, making now an ideal time to make predictions
as to the statistics of high redshift quasars.  

In this paper we use a new large-scale hydrodynamic cosmological
simulation %\textbf{(REF?)}
to study the earliest supermassive black holes, focusing on their luminosity
and clustering properties.  We take advantage of this simulation's large volume
(it is the largest cosmological simulation which
directly models the growth and evolution of black holes) to investigate the
statistical properties of the earliest supermassive black holes, focusing on
the luminosity function and correlation length.  We use these quantities to
compare with current observational measurements, to make predictions for
upcoming surveys, and to investigate the implications of potential features which
may be detected. 

In Section \ref{sec:Method} we describe the numerical sumulation used for our analysis.  In
Section \ref{sec:Results} we investigate the Quasar Luminosity Function
(Section \ref{sec:QLF}) and quasar clustering (Section \ref{sec:Clustering})
and compare with observations, and our results are summarized in Section \ref{sec:Conclusions}.

\section{Method}
\label{sec:Method}

In this paper we use a new  cosmological hydrodynamic
simulation of a $533 \: h^{-1}$ Mpc box specifically
intended for high-redshift investigations (see Table \ref{simparam}).  The simulation uses the massively
parallel cosmolocial TreePM-SPH code GADGET-3 \citep[an updated version of
  GADGET-2, see][]{2005MNRAS.364.1105S} incorporating a multi-phase ISM model
with star formation \citep{SpringelHernquist2003} and black hole accretion and feedback
\citep{SpringelFeedback2005, DiMatteo2005}.  

Within our simulation, black holes are modeled as collisionless sink particles
which form in newly emerging and resolved dark matter halos.  These halos are
found by calling a friends of friends group finder at regular intervals (in
time intervals spaced by $\Delta \log a = \log 1.25$). %\textbf{(Confirm the
				%same interval is used)}  
Any group above a threshold mass of $5 \times 10^{10}
h^{-1} M_\odot$ not already containing a black hole is provided one by
converting its densest particle to a sink particle with a seed mass of
$M_{\rm{BH,seed}} = 5 \times 10^5 h^{-1} M_\odot$.  This seeding prescription
is chosen to reasonably match the expected formation of supermassive black
holes by gas directly collapsing to BHs with $M_{\rm{BH}} \sim
M_{\rm{seed}}$ \citep[e.g.][]{BrommLoeb2003, Begelman2006} or by PopIII stars
collapsing to $\sim 10^2 M_\odot$ BHs at $z \sim 30$ \citep{Bromm2004, Yoshida2006} followed by sufficient
exponential growth to reach $M_{\rm{seed}}$ by the time the host halo reaches
$\sim 10^{10} M_\odot$.  Following insertion, BHs grow in mass by accretion of
surrounding gas and by merging with other black holes.  Gas is accreted
according to $\dot{M}_{\rm BH} = \alpha \frac
  {4 \pi G^2 M_{\rm BH}^2 \rho}{(c_s^2 + v^2)^{3/2}}$
  \citep{1939PCPS...35..405H, 1944MNRAS.104..273B, 1952MNRAS.112..195B}, where
  $\rho$ is the local gas density, $c_s$ is the local sound speed, $v$ is
  the velocity of the BH relative to the surrounding gas, and $\alpha$ is introduced to correct for the reduction of the gas density close
  to the BH due to our effective sub-resolution model for the ISM. To allow for the
  initial rapid BH growth necessary to produce sufficiently massive BHs at
  early time ($\sim 10^9 M_\odot$ by $z \sim 6$) we allow for mildly
  super-Eddington accretion \citep[consistent with][]{VolonteriRees2006,
    Begelman2006}, but limit it to a maximum of $3 \times \dot{M}_{\rm{Edd}}$
  to prevent artificially high values.

\begin{table}
\caption{Numerical Parameters}
\begin{tabular}{c c c c c}

  \hline
  \hline

Boxsize & $N_p$ & $m_{DM}$ & $m_{gas}$ & $\epsilon$ \\
$h^{-1}$ Mpc & & $h^{-1} M_\odot$ &  $h^{-1} M_\odot$ &  $h^{-1}$ kpc \\

\hline

533.33 & $2 \times 3200^3$ & $2.8 \times 10^8 M_\odot$ & $5.7 \times 10^7
M_\odot$ & 5.0 \\

\hline

\end{tabular}
\label{simparam}
\end{table}

The BH is assumed to radiate with a bolometric luminosity proportional to the
accretion rate, $L = \eta \dot{M}_{\rm{BH}} c^2$ \citep{ShakuraSunyaev1973},
where the radiative efficiency $\eta$ is fixed to 0.1 throughout the
simulation and our analysis.  To model the expected coupling between the
liberated radiation and the surrounding gas, 5 per cent of the luminosity is
isotropically deposited to the local black hole kernel as thermal energy.  The
5 per cent value for the coupling factor is based on galaxy merger simulations
such that the normalization of the $M_{\rm{BH}}-\sigma$ relation is reproduced
\citep{DiMatteo2005}.  

The second mode of black hole growth is through mergers which occur when dark
matter halos merge into a single halo, such that their black holes fall toward the
center of the new halo, eventually merging with one another.  In cosmological
volumes, it is not possible to directly model the physics of the infalling BHs
at the smallest scales, so a sub-resolution model is used.  Since the mergers
typically occur at the center of a galaxy (i.e. a gas-rich environment), we assume the final coalescence
will be rapid \citep{MakinoFunato2004, Escala2004, Mayer2007}, so we merge the
BHs once they are within the spatial resolution of the simulation.  However,
to prevent merging of BHs which are rapidly passing one another, mergers are
prevented if the BHs' velocity relative to one another is too high (comparable
to the local sound speed).  

The model used for black
hole creation, accretion and feedback has been investigated and discussed in
\citet{Sijacki2007, DiMatteo2008, Colberg2008, Croft2009, Sijacki2009,
    DeGraf2010, DeGrafClustering2010}, finding it does a good job reproducing
the $M_{\rm{BH}}-\sigma$ relation, the total black hole mass density
\citep{DiMatteo2008}, the QLF \citep{DeGraf2010}, and the expected black hole
clustering behavior \citep{DeGrafClustering2010}.  This simple model thus appears
to model the growth, activity, and evolution of supermassive black holes in
a cosmological context surprisingly well (though the detailed treatement of the accretion
physics is infeasible for cosmological scale simulations).  We also note that
\citet{BoothSchaye2009} and \citet{Johansson2008} have adopted a very similar model,
and have independently investigated the parameter space of the reference model
of \citet{DiMatteo2008}, as well as varying some of the underlying prescriptions.  For further details on the simulation methods and
convergence studies done for similar simulations, see \citet{DiMatteo2008}.

Because the simulation saves the complete set of black hole properties (mass,
accretion rate, position, local gas density, sound speed, velocity, and BH
velocity relative to local gas) for each BH at every timestep, the black hole output for
such a large simulation is prohibatively difficult to analyze using previous
techniques.  For this reason, \citet{Lopez2011} developed a relational database
management system specifically for this simulation.  A similar strategy has
also been followed in the analysis of the Millenium simulation
\citep{Lemson2006}.  In
addition to providing a substantially more efficient query system for
extracting information, this database is significantly more flexible than
traditional approaches.  For a complete summary of the database format and its efficiency,
please see \citet{Lopez2011}.  

\section{Results}
\label{sec:Results}

\subsection{Luminosity Function}
\label{sec:QLF}

\begin{figure*}
\centering
\subfigure{
\includegraphics[width=8.5cm]{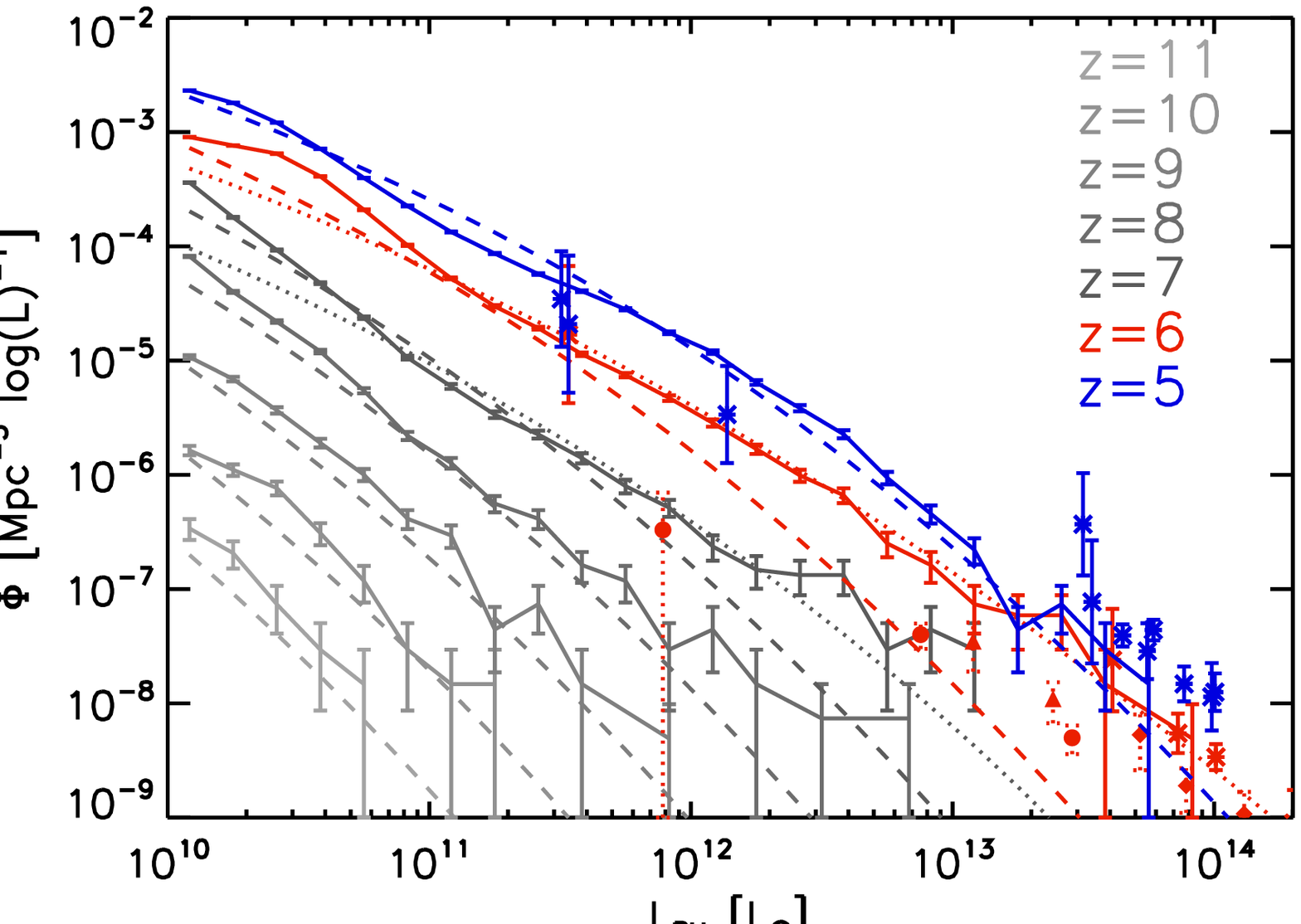}
}
\subfigure{
\includegraphics[width=8.5cm]{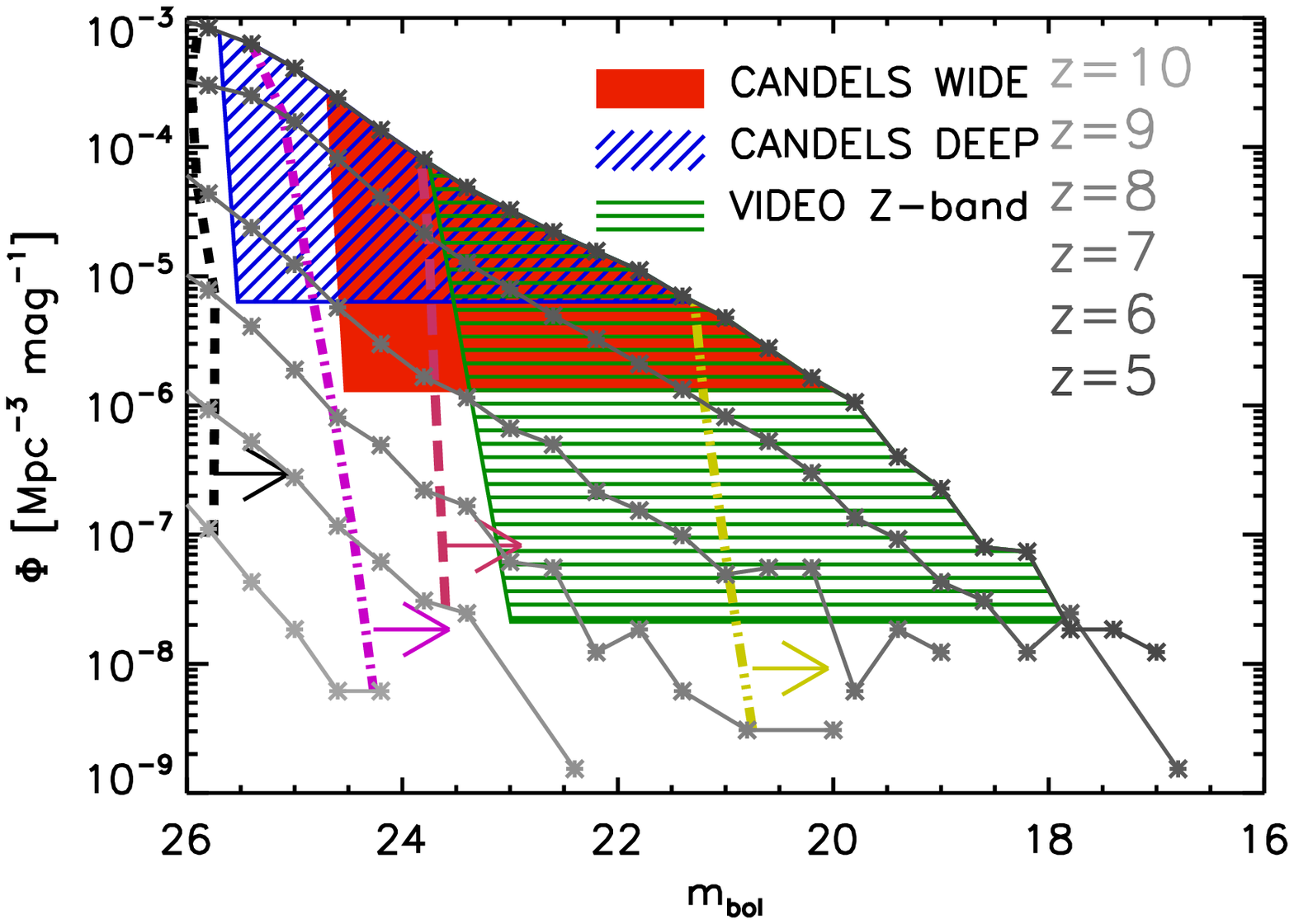}
}
\caption{\textit{Left panel:} Quasar luminosity function for $z=11$ to $z=5$
  (solid lines with Poisson error bars), along with
  the $z=5$ and 6 observational datapoints compiled by \citet{Willott2010} and
  \citet{Hopkins2007} [See text for complete references to observational data].  We also show the
  predicted QLF from our HOD model using the $z=5$ fit (dashed lines) and using
  redshift-specific fits (dotted lines).  [See Table \ref{HODparam} for
  fitted parameters.]  \textit{Right panel:} Quasar magnitude function for $z=11$ to $z=5$,
  based on apparent bolometric magnitude.  The regions probed by CANDELS and
  VIDEO surveys are shown as shaded regions (solid red - CANDELS DEEP; hatched blue -
  CANDELS WIDE; hatched green - VIDEO).  The lower limits are based on the
  survey volume for $z \pm 0.5$.  We also show the magnitude limits for JWST
  NIRCam (black dashed), LSST (pink dot-dashed), Dark Energy Survey (red long
  dashed), and VIKING (yellow triple-dot dash).}
\label{QLF}
\end{figure*}

In the left panel of Figure \ref{QLF} we show the bolometric Quasar Luminosity Function for
$L_{\rm{BH}} > 10^{10} L_\odot$ at $z=11$ to $z=5$ (solid lines).  We also show the
observational data compiled by \citet{Willott2010} from SDSS main
\citep[diamonds,][]{Fan2006}, SDSS deep stripe
\citep[triangles,][]{Jiang2009}, and the Canda-France High-z Quasar Survey
\citep[circles,][]{WillottCFHQS2010}, and that compiled by
\citet{Hopkins2007} \citep[asterisks;][]{Fan2001, Fan2001b, Barger2003,
  Barger2003b, Fan2003, Fan2004, Cristiani2004, Barger2005, Richards2005, Silverman2005}.  %\textbf{(Should all surveys compiled by Hopkins be cited here?)}  
We find that our simulation is generally consistent with
observations, though we tend to predict a steeper slope than
observations.  Thus this large simulation volume shows that our model is
capable of producing Sloan-type BHs at high redshift ($z=5,6$), and in the correct abundances
\citep[and in fact of sufficiently high mass, see][]{DiMatteo2011}.  

To help better understand the quasar populations which we are simulating and
the dark matter halos in which they are found, we have
performed a simple fit to characterize the relation between BH luminosity and
host halo mass.  To do this, we use a simple Halo
Occupation Distribution (HOD) model.  Because we are only interested in the
high-luminosity sources (and it is exceedingly rare to find multiple
high-luminosity objects within a single halo, especially at such high redshift), we model
the probability that a halo of a given mass ($M_{\rm{Host}}$) will host a
BH above a specified luminosity ($L_{\rm{BH,cut}}$) as a cumulative log-normal
distribution \citep[which is often used in galaxy HOD models for $\langle
  N_{\rm{cen}} \rangle$, see, e.g.][]{Zheng2005}

\begin{equation}
  f(L_{\rm{BH,cut}}|M_{\rm{Host}}) = \frac{1}{2} \left ( 1+ \rm{erf} \left (
  \frac{\ln(M_{\rm{Host}})-\mu (L_{\rm{BH,cut}})}{\sqrt{2 \sigma^2 (L_{\rm{BH,cut}})}} \right )
  \right )
\label{eqn:occupationfit}
\end{equation}
where $f(L_{\rm{BH,cut}}|M_{\rm{Host}})$ is the fraction of halos with mass
$M_{\rm{Host}}$ which contain a BH above $L_{\rm{BH,cut}}$, and erf is the
error function $\left ( \rm{erf}(x)=\frac{2}{\sqrt{\pi}} \int_0^x e^{-y^2}dy
\right )$ [see Table \ref{HODparam} for fitted parameters].  We combine this
occupation fraction (using parameters at $z=5$) with the halo mass function of
\citet{Reed2007} to predict the
QLF, shown in Figure \ref{QLF} (blue dashed line).  This predicted
QLF matches the simulation well at $z=5$, confirming an
accurate fit.  We note that our HOD approach may slightly underpredict the low
luminosities since the HOD parameters are fit for bright ($> 10^{11} L_\odot$)
quasars, and because at low luminosities, satellite BHs (which are not
included in our simple HOD) may become significant.  For bright objects,
however, this approach is very accurate.  

We also plot the QLF for higher redshifts using the $z=5$ HOD fit (dashed lines), and find that
it does a surprisingly good job of predicting the higher redshift QLF,
with the expection of the luminous BHs at $z=6$ and $7$.  This suggests
that the manner in which BHs populate halos is approximately redshift
independent, save for the luminous BHs at redshifts 6 and 7.  At
these redshifts, we find the rarest (i.e. brightest) objects to be brighter than our
$z=5$ fit would expect.  In other words, at $z=6$ and $7$, the BHs in the most
massive halos are more luminous than those in comparable halos at other
redshifts, a result of unusually high accretion rates.  To quantify this
discrepancy, we performed HOD fits for $z=6,7$ (Table \ref{HODparam}) which
are shown as dotted lines in Figure \ref{QLF}.
We note that $A_\mu$ remains approximately constant, suggesting that BHs of
$\sim 10^{11} L_\odot$ tend to populate similar-size halos regardless of
redshift (at least for $z \ge 5$).  However, at $z=6$ and $7$ the smaller $B_\mu$
implies that
BHs found in massive halos at $z=6$ and $7$ tend to be brighter than
those in comparable halos at other redshifts, and the effect gets stronger for
higher mass halos/higher luminosity BHs.  

This can be explained by considering the growth history of the massive BHs.
\citet{DiMatteo2011} find that the most massive BHs typically undergo a
period of rapid (i.e. Eddington) growth in this general redshift range ($z \sim 6-8$)
as a result of high density streams of cold gas which not only help assemble
the first massive halos, but appear to penetrate to the halo centers.  These
cold flows facilitate extended periods of Eddington growth of BHs in the most massive
halos, which manifests
itself in the QLF by flattening the bright end.  This continues until
the energy released by the quasar heats the surrounding gas to such a point
that it is blown away from the halo center, and the BHs become
self-regulated \citep[see][for an investigation into this
  behavior]{DiMatteo2011}, at which point the bright-end will steepen once again.  In this
picture, the QLF should roughly follow the evolution of the halo mass
function except for those BHs undergoing their Eddington growth phase, where the luminosity should be unusually high.  We
find exactly these results in Figure \ref{QLF}, where the $z=5$ fit does a good
job except in the bright-end at $z=6-7$, where Eddington growth is common.  
%We
%also note that there appears to be an increase in $A_\sigma$ (the spread of the
%$L_{\rm{BH}}$-$M_{\rm{Host}}$ relation), but this may simply be a result of the worse statistics at high redshift.  

In addition to the luminosity function shown in the left panel of Figure \ref{QLF}, we show the magnitude function in terms of apparent bolometric
magnitude in the right panel which we use as a basis for predictions for
several upcoming surveys.  In particular, we show the
regimes probed by the Cosmic Assembly Near-IR Deep Extragalactic Legacy Survey
(CANDELS) WIDE and DEEP surveys\footnote{\url{http://candels.ucolick.org}} (solid red
and hatched blue, respectively).  These regions are bounded by the magnitude
limit of the F125W filter \citep[converted to bolometric magnitude using the spectral
  energy distribution (SED) of][]{Hopkins2007}, and the volume enclosed by the
survey areas ($m<26.4$ over 0.2 deg$^2$ for WIDE and $m<27.4$ over 0.04 deg$^2$ for DEEP, assuming a redshift range of $\pm 0.5$).  We predict that both the WIDE and DEEP
programs will find sources out to $z \sim 7-8$, with each probing a slightly different
region of the luminosity function.  We expect there to be $\sim 119 \pm 28$ ($\sim 87 \pm 28$) quasars
bright enough in the F125W band of the CANDELS WIDE (DEEP)
fields from $z=5-6$, $\sim 19 \pm 7$ ($\sim 18 \pm 9$) from
$z=6-7$, and $\sim 1.7 \pm 1.5$ ($\sim 1.5 \pm 1.5$) from $z=7-8$.
This will drastically increase the number of high-redshift quasars, which will
substantially improve the measurements of the faint end of the $z=6$
QLF, and the different (though overlapping) ranges probed by the WIDE and DEEP
programs will help constrain the faint-end slope.  We have also provided a
similar bounded volume for the VISTA Deep Extragalactic Observations Survey
(VIDEO), and magnitude limits for
JWST's Near-Infrared Camera (NIRCam; black dashed line) \citep{JWST2006}, Large
Synoptic Survey Telescope (LSST; pink dot-dashed), Dark Energy Survey (red
long dashed), and VISTA Kilo-Degree IR Galaxy Survey (VIKING; yellow
triple-dot dash), each
converted to bolometric magnitudes \citep[using the SED of ][]{Hopkins2007} to
provide the means for predicting the number of high-z quasars these surveys
should find.  Overall, these surveys should drastically improve the
measurements of the faint-end QLF by improving the number of observed quasars,
and pushing to luminosities an order of magnitude lower than current
observations.

\begin{figure}
\centering
\includegraphics[width=9cm]{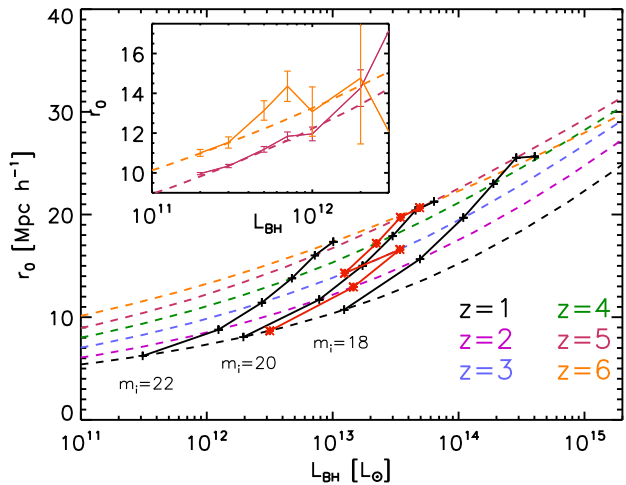}%plots/3200_xi_actual_HODprediction.ps}
\caption{\textit{Main panel:} The predicted correlation length based on our HOD model as a function of luminosity (dashed lines).  Curves
  of constant apparent i-band magnitude are shown in black, and the magnitude
  cut used for quasars by SDSS (19.1 for $z < 3$, 20.2 for $z > 3$) is shown
  in red. Note that the $z < 5$ curves use the $z=5$ HOD fit.  \textit{Inset:}
  Comparison of $r_0$ from the HOD-based prediction (dashed
  lines) and that obtained directly from our simulation (solid lines) for
  $z=5$ (red) and 6 (orange).  }
\label{corr_lum_relation}
\end{figure}

\subsection{Quasar Clustering}
\label{sec:Clustering}

In addition to the number density of quasars, we look at the clustering
properties characterized by the correlation length ($r_0$, defined as the scale
at which the correlation function $\xi (r_0) = 1$).  In Figure \ref{corr_lum_relation} we show the predicted correlation length
as a function of lower-luminosity cut for $z=1$ to 6.  These predicted
correlation lengths are obtained using a halo mass function \citep{Reed2007} and the
halo bias factor of \citet{SMT2001}, weighted by the fraction of halos
hosting sufficiently luminous quasars (Eqn. \ref{eqn:occupationfit}, using the
$z=5$ parameters for $z \le 5$ and the $z=6$ fit at $z=6$).  We note
that the halo bias factor is not well constrained for the very high-mass end \citep[see, e.g.][]{Tinker2010, Pillepich2010},
and has not been investigated significantly at these high redshifts.  
Nonetheless, we
find that the \citet{SMT2001} bias factor does a good job of modeling our halo
bias, though at $z \sim 6$ it underestimates by $\sim 5\%$ relative to our
simulation (note that we adjustment the bias factor by $5\%$ to account for
this when calculating $r_{0,\rm{BH}}$).  This technique is important as it allows us to extend
our predictions to higher luminosities where we do not have sufficient BHs to
obtain clustering statistics.  

To confirm the validity of this approach, we
calculate the correlation length directly from the simulation using a
maximum likelihood estimator \citep[see, e.g.][]{Croft1997}, and show the comparison in the
inset plot (dashed lines - HOD model; solid lines - direct from simulation),
finding good agreement in the luminosity range probed directly by our
simulation.  Thus we confirm that the large-scale clustering of luminous quasars can be
well-modeled by the clustering of dark matter halos and a relatively simple BH
HOD (Eqn. \ref{eqn:occupationfit}).  In the main plot we also show curves of constant apparent magnitude
(solid black curves), and a curve for the magnitude cut used for quasar
selection in SDSS ($m_i < 20.2$ for $z>3$, $m_i < 19.1$ for $z<3$, solid red
curve).  These curves
show that using a magnitude cut rather than a luminosity cut will increase the
observed redshift evolution.  We also see that the evolution of the clustering strength
with luminosity is relatively minor
for low-$L_{\rm{BH}}$ (where BHs occupy the slowly-evolving end of the halo
mass function), but becomes more significant at high luminosities where
BHs occupy the high-mass tail of the mass function.  

In addition, we note the effect of the change in the HOD model from $z=5$
(red) to $z=6$ (orange).  For low luminosities (below $\sim 10^{13} L_\odot$)
there is minimal effect, due to the similarity in $A_\mu$.  However, the
smaller $B_\mu$ at $z=6$ (Table \ref{HODparam}) means that higher
luminosity BHs will typically be found in smaller halos, which have a
correspondingly lower correlation length.  Because the effect grows with
luminosity, the difference manifests itself in the
shallower slope of $r_0$ vs. $\rm{L}_{\rm{BH}}$, even resulting in a crossover at
$\sim 4 \times 10^{13} L_\odot$.  We note that this crossover is highly
dependent on the HOD at high luminosities, and we are forced to extrapolate
our HOD to reach these values.  Thus this crossover is not a strong prediction (see
further discussion below) but the general suppression is a clear trend in our
simulation.  In this way we see that $r_0$ could be effected by a change in
the HOD, but the effect should be fairly minor and will only occur at extremely
high luminosities ($\simgt 10^{13.5} L_\odot$).

\begin{figure}
\centering
\includegraphics[width=9cm]{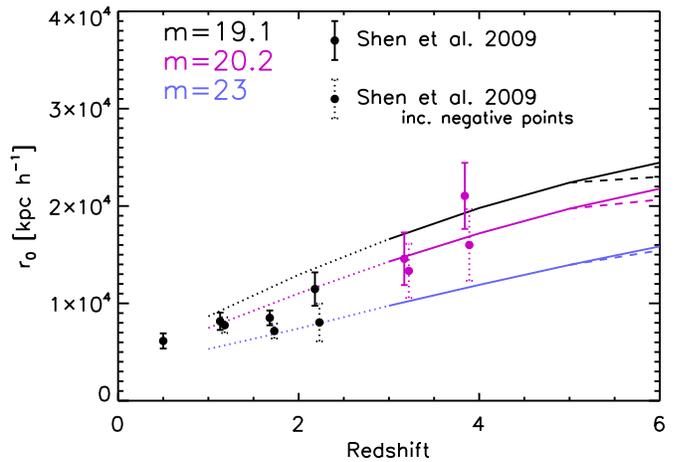}%3200_correlationlength_HODprediction.ps}
\caption{Evolution in the predicted correlation length assuming a redshift
  independent HOD fit (using $z=5$ parameters) for magnitude
  cuts of 19.1 (black, SDSS QSO cut for $z<3$), 20.2 (pink, SDSS QSO cut for $z>3$), and 23.0 (blue).  The dashed line from
  $z=5$ to 6 shows $r_0$ if $z=6$ HOD fit is
  used at $z=6$.  \textit{Filled circles:}
  Observational data from
  \citet{Shen2009} obtained with and without negative datapoins (dashed and solid error
  bars, respectively).  }
\label{correlationlength}
\end{figure}

In Figure \ref{correlationlength} we show the evolution in our predicted correlation length
for SDSS-type quasars based on our $z=5$ HOD fit (for luminosity cuts
corresponding to $m_i <19.1$ - Black; $m_i < 20.2$ -
Pink; $m_i <23.0$ - Blue), together with the observed measurements of \citet{Shen2009} (obtained
with and without inclusion of negative datapoints, shown with dashed and solid
error bars respectively).  The curves assume constant occupation behavior (i.e. the $z=5$
HOD parameters are used at all redshifts).  Based on earlier work, we would expect
relatively minor evolution in the HOD between $z=5$ and $z=3$
\citep{Chatterjee2011}, so the extrapolation should remain fairly accurate for
$z > 3$.  This extrapolation shows that our simulation is able to reproduce
the large correlation lengths ($r_0 \sim 17  \: h^{-1} \: \rm{Mpc}$ at $z=4$)
from observational constraints.  In addition, we match the rough evolution
with redshift for $z>3$, and we expect $r_0$ to continue to increase with redshift, reaching $\sim 22 \: h^{-1} \: \rm{Mpc}$ by $z=6$.
Because we use a redshift-independent BH HOD model, this agreement with the
observed evolution of $r_0$ with redshift suggests that high-z quasar clustering
evolution can be completely explained by the evolution in the clustering of
dark matter halos, i.e. without evolution in the mass of the typical host halo.   
Below $z=3$ we expect the AGN HOD to evolve more quickly
\citep{Chatterjee2011}, so the extrapolation should not be considered an
accurate prediction at lower redshifts (the dotted curves of Fig. \ref{correlationlength}), but we nonetheless appear to remain broadly consistent with observations.  

In addition to the curves for the $z=5$ HOD, we also show the effect of the
change in occupation distribution from $z=5$ to 6 (dashed lines).  At $z=6$
the shallower slope in the $L_{\rm{BH}}-M_{\rm{Host}}$ relation means that a
given luminosity cut will correspond to a smaller typical host mass (and thereby
produce a smaller $r_0$) at $z=6$ than $z=5$.  
Because the difference in the HOD is in the slope ($B_\mu$) of the $L_{\rm{BH}}-M_{\rm{Host}}$ relation,
the suppression gets stronger at higher luminosities and has essentially no
effect on lower luminosities, which we show with
the m$_i=23.0$ curve of Figure \ref{correlationlength}.  We note that the
magnitude of the suppression is highly dependent on the slope of $\mu$ in
Equation \ref{eqn:occupationfit} (found in Table \ref{HODparam}) and is only
significant at luminosities above those well-probed by our simulation.  As
such, we cannot accurately estimate the magnitude of the suppression (or the
exact luminosity at which the $z=5$ and 6 curves of Figure \ref{corr_lum_relation} will cross), but we
nonetheless expect a suppressed slope in the $r_0$ evolution from $z=5$ to 6.
Given the approximate magnitude of the suppression found in our simulation (on
the order of $\sim5\%$) and
the difficulties in measuring $r_0$ to high precision at such high redshifts, it is
unlikely that such suppression will be detected in the near future, but it
is nonetheless an effect which should be considered wherever possible.  

%\textbf{Should there be any additional discussion of evolution to lower z? i.e. the
%  evolution to lower z will explain the shape of the observational data (BHs
%  being typically fainter will mean a given L will be found in larger halos,
%  and therefore the evolution with z will be suppressed), but that will make
%  our actual values agree worse.  Also, the bias factor may continue to evolve
%  as we do that, which may also be a factor, but is this worth discussing?.  }

\begin{table}
\caption{Luminosity dependence of $\mu$ and $\sigma$ (see Eqn.
  \ref{eqn:occupationfit}). \newline $\mu,\sigma(L_{\rm{BH,cut}}) = A_{\mu,\sigma} \times \left (\frac{L_{\rm{BH,cut}}}{10^{11} L_\odot} \right )^{B_{\mu,\sigma}}$}
\begin{tabular}{c c c c c}

  \hline
  \hline
 
  Redshift & $A_\mu$ & $B_\mu$ & $A_\sigma$ & $B_\sigma$ \\

\hline

  5 & 27.05 & .0208 & .209 & .128 \\
  6 & 26.905 & .0191 & .239 & .133 \\
  7 & 27.07 & .0173 & .298 & 0 \\
%  5 & 26.975 & .0229 & .206 & .1415 \\
%  6 & 26.912 & .0188 & .239 & .11186 \\
%  7 & 27.063 & .0175 & .268 & .156 \\

\hline

\end{tabular}
\label{HODparam}
\end{table}

\section{Conclusions}
\label{sec:Conclusions}

In this paper we have investigated high redshift black holes found within a
new large-scale hydrodynamic simulation, focusing on the luminosity function
and clustering behavior.  We model the QLF for $z \ge 5$, probing black holes
to luminosities up to $\sim 10^{14} L_\odot$.  We find reasonable agreement with
observational data at $z=5$ and 6, generally falling within the variation
%\textbf{Better term than variation?}
 between surveys, confirming that our model is capable of producing Sloan-type
 quasars at very early times as well as matching their observed abundances.  

Using a HOD fit for
the central AGN at $z=5$ (which we have provided), we find that the evolution in the QLF is
well described by the evolution in the halo mass function (at least for
the redshifts modeled by our simulation) without significant evolution in the
occupation distribution, except for the high-luminosity end at $z=6-7$ where we find a significantly flatter luminosity function.  We
postulate this flattening of the luminosity function to be a result of the
largest black holes tending to undergo unusually rapid growth (and thereby
producing unusually high luminosities) during these redshifts, a conclusion
supported both by the difference in the HOD fits at these redshifts,
and by direct investigation into the lightcurves of these black holes
\citep{DiMatteo2011}.  At lower redshift (by $z \sim 5$), self-regulation
suppresses the most massive BH growth, resulting in fainter bright-end BHs,
and a corresponding steepening of the QLF.  In particular, we note that this
increase in luminosity at $z \sim 6-7$ should make
observations of such high-z, luminous ($>10^{12} L_\odot$) BHs easier than
would otherwise be expected.  

We used our luminosity function to provide estimates on the number density of
detectable high redshift quasars for several upcoming surveys.  In particular,
we expect the CANDELS survey to find quasars up to $z \sim 7$ in both the WIDE and
DEEP programs, with each probing a slightly  different region of the QLF.  At $z
\sim 6$ these programs should each detect dozens of quasars across a wide range of luminosities, drastically improving the observational constraints on the faint end of
the high redshift QLF, particularly the faint-end slope.  In addition, we have provided our estimated number
density at each redshift for several additional upcoming surveys, thereby providing
the approximate number density of quasars to be found as well as the survey
areas necessary to reach the highest
redshifts.

We also investigate the clustering behavior of these high redshift quasars,
finding luminosity dependent correlation lengths on the order of  $\sim 10-14
\: h^{-1} \rm{Mpc}$.  Using our HOD model and the theoretical clustering of
dark matter halos, we show the correlation length as a function of BH
luminosity, finding relatively weak luminosity dependence at low
$L_{\rm{BH}}$, but with increasing $L_{\rm{BH}}$-dependence at higher
luminosities (where black holes are typically found in halos in the steep
end of the halo mass function).  We also compare our predicted $r_0$ to
high-redshift observations and find excellent agreement, with our simulation
predicting $r_0 \sim 15-20 \: h^{-1} \: \rm{Mpc}$ for Sloan-type quasars.  We
also roughly match the evolution of $r_0$ with redshift using a
redshift-independent HOD, suggesting that high-redshift quasar clustering
evolution is fully explained solely by the evolution in clustering of dark
matter halos without a change in the typical host halo mass.  

Finally, we note the effect a change in the BH occupation distribution
from $z=5$ to 6 has on $r_0$, finding that it should suppress the clustering
strength of high luminosity quasars at $z=6$.  Our limited sample
size and evolution in the halo bias factor found in our simulation make it
difficult to quantify the magnitude of this suppression, but we nonetheless
conclude that some suppression should
occur (though likely below the sensitivity of upcoming surveys).

\section*{Acknowledgments}
%\textbf{Be sure to thank Julio.  Get list of grants used.}
This work was supported by the National Science Foundation, NSF Petapps,
OCI-0749212 and NSF AST-1009781.  The simulations were carried out on Kraken
at the National Institute for Computational Sciences (http://www.nics.tennessee.edu/).

 \bibliographystyle{mn2e}	% or "unsrt", "alpha", "abbrv", etc.
 \bibliography{astrobibl}	% use data in file "astrobibl.bib"

\end{document}